\documentclass[useAMS,usenatbib,a4]{mn2e}

\usepackage{graphics}
\usepackage{graphicx}
\usepackage{natbib}
\usepackage{paralist}
\usepackage{subfigure}
\usepackage{amssymb}

\title[GSH~006$-$15+7: Supershell transitioning from emission to absorption]
{GSH~006$-$15+7: A local Galactic supershell featuring transition from H{\sc i} emission to absorption}
\author[V.A. Moss, N.M. McClure-Griffiths et al.]
{V.A. Moss$^{1,2}$, N.M. McClure-Griffiths$^2$, R. Braun$^2$, A.S. Hill$^2$ and G.J. Madsen$^1$
\\
$^1$ Sydney Institute for Astronomy, School of Physics A29, The University of Sydney, NSW 2006, Australia \\
$^2$ CSIRO Astronomy and Space Science, ATNF, PO~Box~76, Epping NSW~1710, Australia.}

\let\leqslant=\leq %Authors with AMS fonts and mssymb.tex can comment
\date{}

\begin{document}
\maketitle
\newcommand{\setthebls}{
%                 de-comment this line for double spacing:
%\baselineskip=20pt
}
\setthebls
\begin{abstract}
We report on the discovery of a new Galactic supershell,
GSH~006$-$15+7, from the Galactic All Sky Survey data. Observed and
derived properties are presented and we find that GSH~006$-$15+7 is 
one of the nearest physically large supershells known, with 
dimensions of $\sim 780 \times 520$~pc at a distance of $\sim$ 1.5~kpc. 
The shell wall appears in H{\sc i} emission at $b \lesssim -6.5^\circ$
and in H{\sc i} self-absorption (H{\sc i}SA) at $b \gtrsim -6.5^\circ$. 
We use this feature along with H{\sc i}SA diagnostics to 
estimate an optical depth of $\tau~\sim$~3, a spin temperature of~$\sim$~40~K and a swept-up 
mass of M $\sim 3 \times 10^6~\mathrm{M}_\odot$. We also investigate the origin of
GSH~006$-$15+7, assessing the energy contribution of candidate powering sources 
and finding evidence in favour of a formation energy of $\sim 10^{52}$ ergs.
We find that this structure provides evidence for the transfer of mass and energy from 
the Galactic disk into the halo.
\end{abstract}


\section{Introduction}
The pervasiveness of neutral hydrogen (H{\sc i}) in the universe has ensured its centrality in both Galactic and extragalactic studies, from star-forming regions to supernovae to disc-halo interaction. Large shell-like structures detected in H{\sc i}, known as supershells \citep{Heiles:1976p20084}, are an important driving force in the Galactic ecosystem, circulating material within the disc, evacuating dense regions and encouraging new star formation \citep{Cox1974a,McKee1977a,Bruhweiler:1980p15707,McCray:1987p12912,Weisz:2009p20088}. It is theorised that supershells also provide a pivotal link between the disc of a galaxy and the material in its halo \citep{Dove:2000p14867}, indicating that there should be shared properties between halo clouds that are Galactic in origin and supershells on the verge of blowing out of the Galactic disc \citep{Lockman:2002p19226,Ford:2008p9576}. 

Supershells are believed to be formed through the combined effects of multiple supernovae exploding in a region on the order of parsecs to hundreds of parsecs in size, often around an OB association \citep{Heiles:1979p14843}. It is not clear whether this combined effort occurs during one generation of high-mass stars or whether it is the outcome of many generations of stars within a region \citep{Oey:2005p17805}, but the resulting supershell of expanding neutral hydrogen undoubtedly requires significant energy input, from 10$^{52}$ to as much as 10$^{54}$~ergs, in order to become a physical structure hundreds of parsecs in size. The lifetime of a typical supershell is of the order of $10^7$ years \citep{TenorioTagle:1988p13608}, meaning that these structures significantly outlast their parent supernova remnants \citep[$\sim 10^5$ years,][]{Guseinov:1985p8156} and can trace past star formation \citep{Weisz:2009p20088,Anathpindika:2011p20115,Cannon:2011p20097} as well as induce new star formation \citep{Ortega:2009p20104}. It is generally difficult to identify the stars which formed an old supershell, given that most of the high-mass stars will have been extinguished as supernovae in the process of forming the shell \citep{McCray:1987p12912}. If a potential powering source is identified, it is possible to assess whether the formation of the vacuous cavity can be accounted for by its winds and/or supernovae \citep{Brown:1995p15526,Shull:1995p1521,McClureGriffiths:2001p13769,Smith:2006p18022}; in most cases, however, it is impossible to find a source region in an old shell ($t >$ 5 Myr).

We report on the discovery of a Galactic supershell, GSH~006$-$15+7, an object which exhibits a coherent structure over an angular size of $\sim 25^\circ$. In Section \ref{properties}, we describe its observed and derived properties in the context and theory of known Galactic supershells and investigate multiwavelength data in order to characterise GSH~006$-$15+7. In Section \ref{hisa}, we use the unusual property of transition from H{\sc i} emission to self-absorption visible in the walls of the shell to derive spin-temperature, kinetic temperature and optical depth, which allows us to better estimate the mass of the shell. Finally, in Section \ref{model}, we look for potential powering sources that could account for the energy required to form the shell. This supershell presents an interesting case study as one of the nearest large supershells. It highlights the potential dangers of assuming negligible optical depth when estimating mass from column density as well as being a possible example of a supershell in the transition stage between expansion and blowout. Based on the candidate powering sources, we also find evidence in favour of an origin for the supershell in the Sagittarius spiral arm.

\section{GSH~006$-$15+7: Morphology and physical properties}\label{properties}
GSH~006$-$15+7 was discovered within the Galactic All-Sky Survey (GASS) data set, a recent southern-sky H{\sc i} line data survey \citep{McClureGriffiths:2009p3462,Kalberla:2010p19880}. GASS covers the entire sky with declination $\delta \le 1^\circ$ at an angular resolution of $\sim 16$ arcmin, a velocity resolution of $\sim 1$~km~s$^{-1}$ over the velocity range $-400 \le v \le 500$~km~s$^{-1}$ and with an RMS brightness temperature of $\sim 55$ mK. The data set was produced using the 21 cm multibeam receiver on the Parkes 64~m radio telescope. 

\begin{figure}
\includegraphics[angle=0,width=0.5\textwidth]{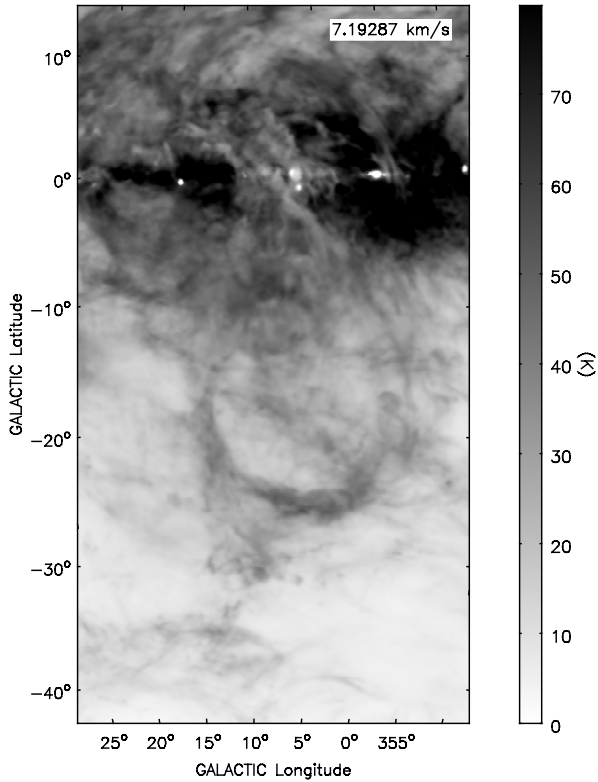}
\caption{H{\sc i} image of GSH~006$-$15+7 at the systemic velocity of 7 km s$^{-1}$. The brightness ranges from 10$-$80 K on a linear~scale. Each velocity slice has a width of $\sim 1$ km s$^{-1}$.}
\label{fg:ib}
\end{figure}

\begin{figure*}
\includegraphics[angle=0,width=0.9\textwidth]{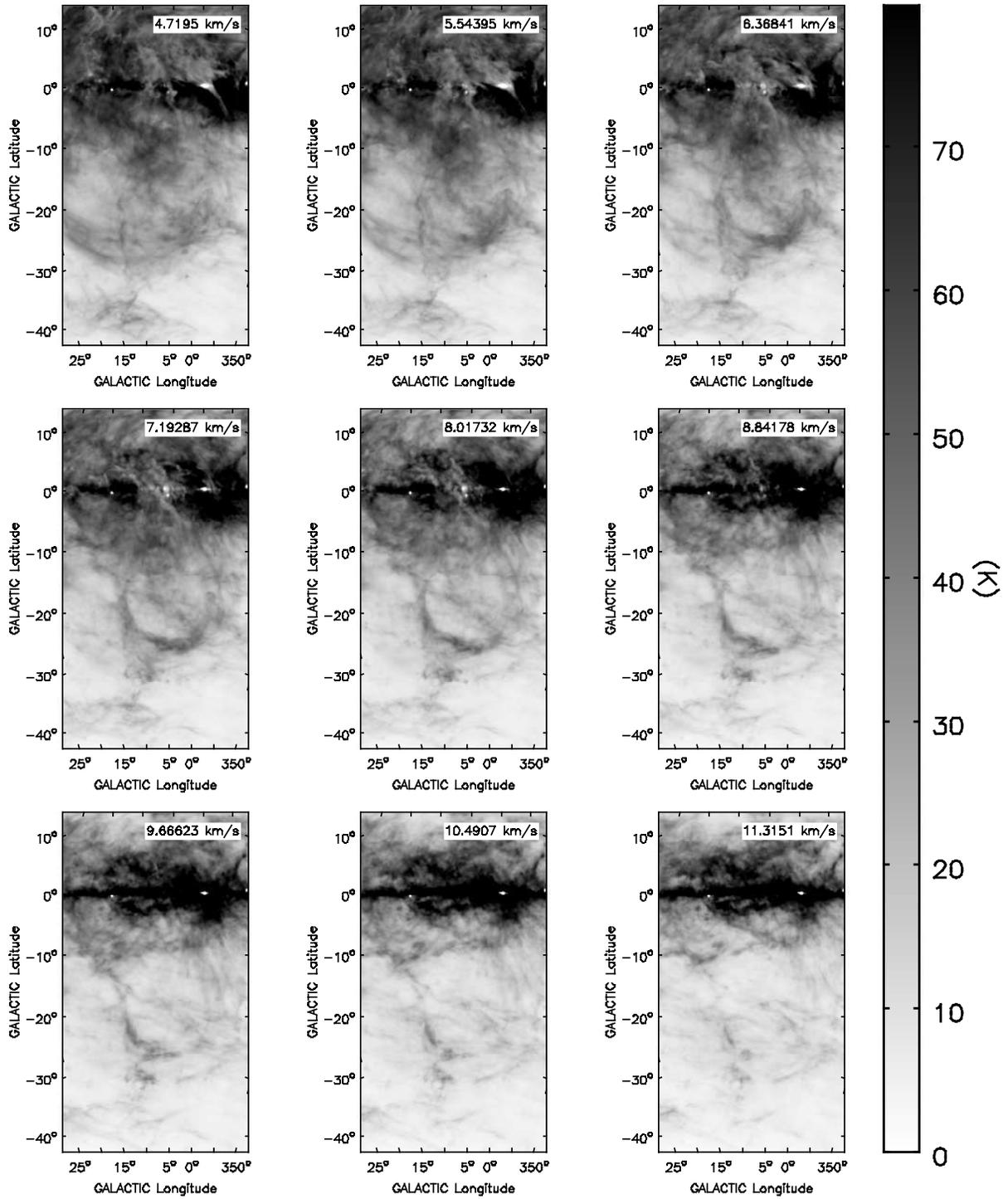}
\caption{H{\sc i} velocity slices of GSH~006$-$15+7, over the velocity range 5$-$11 km s$^{-1}$ in steps of $\sim 1$ km s$^{-1}$. The brightness ranges from 10$-$80 K on a linear scale. Due to overlapping local emission at lower velocities, we see the structure of the shell most clearly at higher velocity although it is fainter.}
\label{fg:ib2}
\end{figure*}

\subsection{General morphology}\label{morph}
The shell is named GSH~006$-$15+7 after its central longitude ($\sim 6^\circ$), central latitude ($\sim -15^\circ$) and systemic velocity\footnote{All velocities are quoted with respect to the Local Standard of Rest (LSR).} (v$_{\rm {LSR}} \sim 7$ km s$^{-1}$). GSH~006$-$15+7 is an example of a well-defined supershell of large angular size, comparable to those described in previous surveys for Galactic supershells \citep{Heiles:1979p14843,Hu:1981p16331,Heiles:1984p12162,McClureGriffiths:2002p12858}. It extends over  longitudes of $l = [356, 16]^\circ$  and latitudes of $b = [-28, 2]^\circ$, and is evident over the velocity range $v_{\rm{LSR}} \sim 5-11$ km s$^{-1}$. Fig. \ref{fg:ib} shows the shell in the velocity slice corresponding to its assumed systemic velocity\footnote{We note that the systemic velocity here is taken to be the apparent central velocity of the shell, which also corresponds to the point of maximum absorption in its spectral profile (Section \ref{hisa}).}, $v_{\rm {LSR}} \sim 7$~km~s$^{-1}$, while Fig. \ref{fg:ib2} shows its morphology over the multiple velocity slices in which it is apparent. We expect that there is emission due to the shell at velocities below 5 km s$^{-1}$ which cannot be discerned due to the confusion of local emission. Table \ref{tb:param} provides a summary of all observed and derived properties.

The shell is teardrop-shaped with its apex crossing through the Galactic plane and extends to a latitude of $-$25$^\circ$.  This teardrop morphology suggests an origin in the Galactic plane, with material elongated in the latitude direction. GSH~006$-$15+7 is an asymmetric structure as there is no convincing evidence of an upper-latitude counterpart with respect to the Galactic plane. The region the shell occupies is confused due to its low velocity, resulting in considerable overlap with local emission, but the shell remains quite well-defined over a large angular size. We examine the structure of the shell at high latitude for evidence of fragmentation and breakout (Section \ref{mw}), and find an apparent break in the shell wall at $l \sim 10^\circ, b \sim -25^\circ$, particularly at the lowest velocities (see Fig. \ref{fg:halpha}). There is also a transition from H{\sc i} emission to absorption on the upper right wall of the shell towards the Galactic plane over $b = [-7.0,-3.0]^\circ$, typically indicative of dense cool gas (Section \ref{hisa}).  

\subsection{Distance and size}
The small velocity $v_{\rm{LSR}}$ of the shell and its longitude close to the Galactic centre make kinematic distance estimates uncertain as rotation curves here are poorly defined with significant velocity crowding. To deal with this uncertainty, we first calculate the kinematic distances and then constrain the results based on knowledge of supershell origins and dynamics. We use a standard Galactic rotation curve \citep{Fich:1989p12134} to estimate the distance to the shell based on the systemic velocity of $\sim 7$~km~s$^{-1}$, namely

\begin{equation}
\frac{\omega}{\omega_0} = 1.00746 \left( \frac{R_0}{R} \right) - 0.017112,
\end{equation}
where $\omega$ is angular velocity, $\omega_0$ is the angular velocity of the Sun, $R_0$ is the Sun's distance from the Galactic centre and $R$ is Galactocentric distance, using $R_0 = 8.3$~kpc and $\Theta_0$ = 246 km s$^{-1}$ in agreement with the most recent literature \citep{Ghez:2008p19997,Gillessen:2009p19985,Bovy:2009p19918}. For an input longitude, latitude and velocity of $l = 6^\circ, b = -15^\circ, v = 7$ km s$^{-1}$ (the central coordinates of the shell and its systemic velocity), the resulting distance is $d \sim 2.0$~kpc for the near distance, and $d \sim 15.5$~kpc for the far distance. We do not further consider the far distance estimates due to the large angular size of the shell and corresponding physical size at such a distance. Due to the large angular scale of GSH 006-15+7, we concentrate on longitude $l = [7, 10]^\circ$ and latitude $b = [-2, 1]^\circ$. This is the region where we might expect to find a powering source based on shell morphology, and results in a near distance range of $d = [1.2, 1.6]$ kpc. Rotation curve distance estimates generally provide a distance that can be off by as much as 2$-$3~kpc due to peculiar motions and local irregularities, an effect which is amplified particularly at low velocities \citep{Gomez:2006p18710,Baba:2009p18881}. If we consider the velocity gradient for the position of the shell over the velocity range [1,~10]~km~s$^{-1}$, this gradient equates to 0.3~kpc~(km s$^{-1}$)$^{-1}$, which indicates the magnitude of the error on the distance calculation. We hence use circumstantial factors to better constrain the distance to GSH~006$-$15+7. 

Looking towards the Galactic centre at longitudes between [0, 10]$^\circ$ on the near side of the Milky Way, we expect to cross the Sagittarius arm (1$-$2 kpc away), the Scutum-Centaurus arm ($\sim 3$ kpc away), the Norma arm ($> 4$ kpc away) and the near 3 kpc arm ($> 5$ kpc away), where distances are estimated based on the best knowledge of Milky Way spiral structure using available data \citep{Churchwell:2009p18522}. We note that there is a significant amount of star formation associated with a distance between 1.2 and 1.8 kpc, in the Sagittarius arm \citep{SteimanCameron:2010p18540}, which agrees well with the kinematic distance of the shell. At a further distance of 3 kpc, if GSH~006$-$15+7 were located on the Scutum-Centaurus arm, we calculate the dimensions of the shell as $\sim$ 1600 $\times $ 1000~pc. This would place the shell in the highest dimensions of known shells, especially for an object observed as a spatially coherent structure without clear signs of blowout. Based on this we believe that the shell is located near the Sagittarius arm and assign it a distance of $1.5\pm0.5$ kpc, with the errors reflecting the range of our kinematic distances. 

\begin{table}
\caption{Basic parameters for GSH~006$-$15+7.}\label{tb:param}
\begin{tabular}{llllllllll}
\hline
Parameter &&& Value\\ 
\hline
Centre ${l}$	& & &	6.0$^\circ$\\
Centre ${b}$	& & &	-15.0$^\circ$\\
Central ${v_{\rm{LSR}}}$	& & &	+7 km s$^{-1}$\\
Distance	& & &	1.5 $ \pm $ 0.5 kpc\\
Dimensions (angular)	& & &	30$^\circ \times$ 20$^\circ$\\
Dimensions (physical)	& & &	$790 \times 520$ pc\\
$v_{\rm{exp}}$		& & &	$\sim$ 8 km s$^{-1}$ \\
$\rm{E_e}$	& & &	$\sim 10^{52}$ ergs\\
Mass	& & &	$3 \pm 2 \times 10^{6}$ M$_\odot$\\
Age	& & &	$15 \pm 5$ Myr\\

\hline
\end{tabular}
\end{table}

At an adopted distance of 1.5 kpc, the shell's angular dimensions of $30^\circ \times 20^\circ$ (latitude extent $\times$ longitude extent) translate into physical dimensions of $790 \times 520 $ pc. Based on these dimensions we can obtain a characteristic radius, $R_{\rm sh}~=~\sqrt{R_{\rm maj} \times R_{\rm min}} \sim$~300$\pm100$~pc. We adopt this characteristic radius in order to estimate the properties of the supershell and to compare it with the known supershell population. A characteristic radius of $\sim 300$ pc places the supershell in the largest 40 per cent  of supershells, where the median shell radius is $\sim$ 210 pc \citep{Heiles:1979p14843,McClureGriffiths:2002p12858}. However, this sample is biased towards small nearby shells due to the difficulty in resolving small shells at any significant distance. If we instead take both distance and radius into account by dividing the shell radius $R_{\rm sh}$ by the shell distance $d_{\rm sh}$,
\begin{equation}\label{rd}
\delta = \frac{R_{\rm sh}}{d_{\rm sh}}
\end{equation}
where $R_{\rm sh}$ and $d_{\rm sh}$ are in parsecs, we see that the median $\delta \sim 0.05$ and that GSH~006$-$15+7 lies within the top 3 per cent of supershells with $\delta = 0.2$, indicating that it is one of the nearest physically large supershells discovered so far. This is notable as the majority of other physically large supershells are quite distant in comparison with GSH~006$-$15+7, most likely due to the difficulty in identifying objects of large angular size. If the shell does extend nearly 800 pc off the Galactic plane as calculated, then we expect that it is almost reaching the expected scale height of Galactic H{\sc i} \citep{Dickey:1990p13348}. 

\subsection{Mass, energy and age}\label{mass}
To estimate the mass of the shell, we first approximate its shape as a cone situated on top of a sphere. Summing the flux over only those velocity slices in which the shell is visible, we produce a zeroth-moment map from which to extract measures of column density (assuming small optical depth), calculated as

\begin{equation}\label{nheq}
N_{\rm H}=  1.823 \times 10^{18} \int_{v_0}^{v_1} T_{\rm b}~dv~\textrm{cm}^{-2},
\end{equation}
where $N_{\rm H}$ is the neutral atomic hydrogen column density, $v_0$$-$$v_1$ represents its spectral extent and $T_{\rm b}$ is the observed brightness temperature. The mean column density was found by averaging the column density measured at different points along the wall of the shell. Assuming an axisymmetric wall width of the shell, we then use the background-corrected column density of the shell. The background was approximated by sampling the column density away from the shell at several points and averaging to find the mean background, avoiding other regions of emission. This gives a mean corrected zeroth-moment of $\sim 117$~K~km~s$^{-1}$ with a standard deviation of $\sim 15$~K~km~s$^{-1}$. We convert the measured size of the cone and sphere from degrees to parsecs assuming a distance of 1.5 kpc and derive a mean column density of $N_{\rm H} \sim 2 \times 10^{20}$~cm$^{-2}$.

Once accounting for the presence of helium (a factor of 1.4), our estimate for mass of the shell comes to $M \sim 10^6~{\rm M_\odot}$ with a number density of $\sim 1.3$~atoms~cm$^{-3}$ in the shell and an ambient density of $\sim 0.7$~atoms~cm$^{-3}$ over the entire volume. Although there is foreground material present, using correlation between H{\sc i} and infrared data shows that the effect of this material on our mass estimate is negligible. We note that our estimate here, typically used to estimate column densities of supershells, is affected by the necessary assumption that the gas is optically thin, and we correct for this assumption in Section \ref{hisa}.

We derive an initial estimate of the total energy of the supershell by calculating the kinetic energy associated with its mass and velocity. We assume the mass to be as estimated ($\sim 10^6~{\rm M_\odot}$). Based on the large physical size of the shell and lack of a clear expansion signature in our data, we assume a late-evolution expansion velocity of 8 km s$^{-1}$, which is generally assigned when shells are in their final phase of expansion and the expansion velocity is comparable to random gas motions \citep{Heiles:1979p14843}. This gives $E_K$~($0.5 M v_{\rm exp}^2$)~$\sim 6 \times 10^{50}$~ergs. If we assume that $E_K$ represents 20 per cent of the total energy input due to efficiency of energy transfer \citep{McCray:1987p12912,Dawson:2008p12949}, then we would expect the total input energy to be $\sim 3 \times 10^{51}$~ergs, which is likely to be an underestimate of the total shell energy based on other supershells of comparable size. We can also obtain a theoretical estimate of the expansion energy required to form the shell as described analytically \citep{Weaver:1977p13671,McCray:1987p12912}, such that

\begin{equation}\label{R}
R = 97~\textrm{pc} (N_*/n_0)^{1/5} t_7^{3/5}
\end{equation}
and 
\begin{equation}\label{vexp}
v_{\rm exp} = 5.7~\textrm{km s}^{-1} (N_*/n_0)^{1/5} t_7^{-2/5}
\end{equation}
where $R$ is the radius in parsecs, $N_*$ is the number of stars with M $>$ 7 M$_\odot$ (including both stars that have gone or will go supernova), $n_0$ is the ambient density in cm$^{-3}$, and $t_7$ is the age in units of 10$^7$~years. To incorporate the level of certainty to which we can specify $n_0$ and $R$, we consider the range $n_0$ = [0.5, 1.0] cm$^{-3}$ (to account for uncertainties consistent with the ambient density predicted above) and $R$ = [200, 400] pc (to account for the asymmetry of the shell). By solving Equations \ref{R} and \ref{vexp} simultaneously, we find that $t_7$ = [15, 30] Myr  and $N_*$ = [6, 47]. This corresponds to an input energy of [1, 10] $\times 10^{52}$ ergs, assuming each star contributes 10$^{51}$~ergs from stellar winds and 10$^{51}$~ergs from going supernova. Overall these energy estimates suggest a formation energy of the order of $10^{52}$ ergs, which is most likely the degree of accuracy to which we can predict the formation energy without further data or modelling. 

For comparison to the overall distribution of supershell energy as performed above for radius \citep{Heiles:1979p14843,McClureGriffiths:2002p12858}, we find a median known supershell energy of $\sim 9 \times 10^{51}$ ergs which places GSH~006$-$15+7 very close to the average formation energy of known supershells, in rough agreement with the percentile of the characteristic radius. Estimating the age of a shell is very difficult in the absence of a current powering source, and so the age represented by $t_7$ is highly uncertain. We can however assume an upper age limit of around 20 Myr, after which the combination of ambient pressure at high scale height, galactic shear and gravitational deceleration likely destroys most coherent structure in a supershell and leads to breakout \citep{TenorioTagle:1988p13608,Mashchenko:1999p16998}. 

\begin{figure}
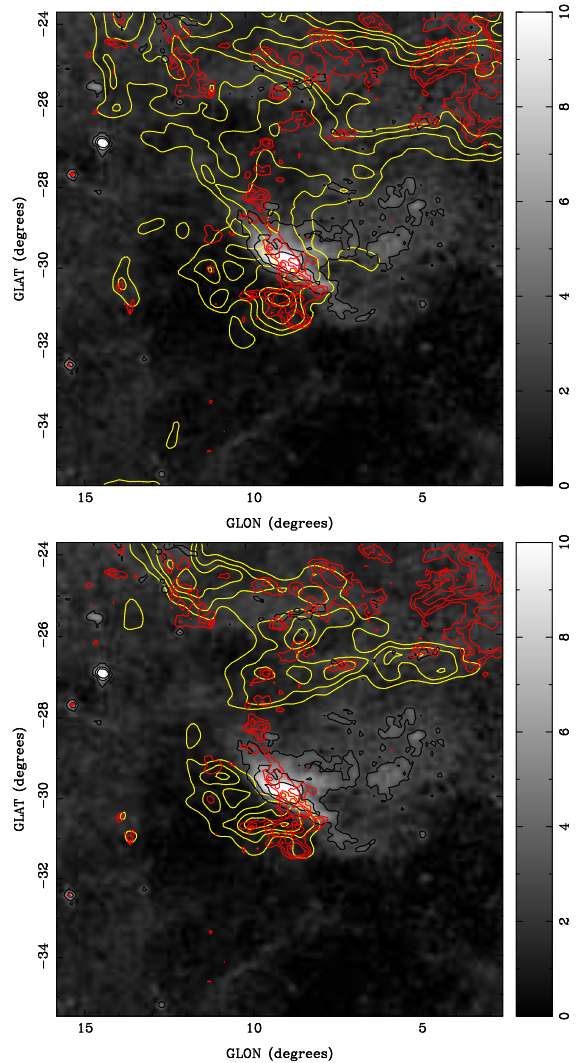

\centering
\subfigure{}
{\includegraphics[angle=-90,width= 0.4\textwidth]{fig3a.eps}}
\subfigure{}
{\includegraphics[angle=-90,width= 0.4\textwidth]{fig3b.eps}}
\caption{Map of H$\alpha$ emission (greyscale) near GSH~006$-$15+7 \citep{Finkbeiner:2003p15462}. In all images, the contours show H{\sc i} emission (yellow, at 25, 30, 35~K) at $\sim$ 7 km s$^{-1}$ (top) and $\sim$ 10 km s$^{-1}$ (bottom), 100 $\mu$m dust (red, at 8, 9, 11~MJy/Sr) and H$\alpha$ emission (black, at 3, 6, 9~R). At 7 km s$^{-1}$, we see strong H{\sc i}/IR correlation.}
\label{fg:halpha}
\end{figure}

\begin{figure}
\centering
\includegraphics[angle=0,width= 0.50\textwidth]{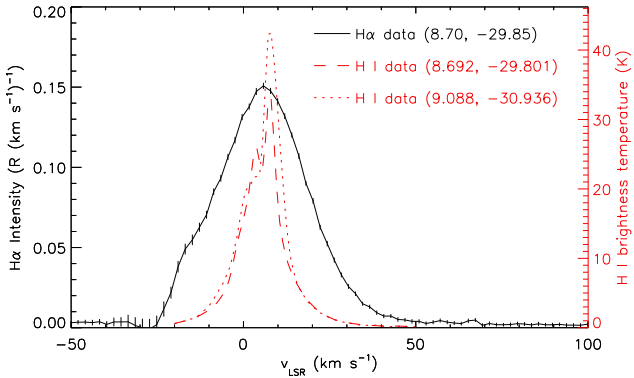}
\caption{H$\alpha$ spectrum obtained with WHAM. This is an on$-$off spectrum, as described in the text. An H{\sc i} spectrum from an average of the GASS beams falling within the WHAM beam is shown with the dashed red line; an average of H{\sc i} spectra from GASS beams falling within a beam the size of the WHAM beam but centred in a nearby knot of bright H{\sc i} is shown with the dotted red line. The geocoronal H$\alpha$ line is incompletely subtracted near $v_{\rm LSR} = -20 \textrm{ km s}^{-1}$.}
\label{halphaspec}
\end{figure}

\subsection{Multiwavelength properties}\label{mw}
We investigated available multiwavelength data to determine if other emission processes are evident. Previously detected emission in supershells include H$\alpha$, soft X-rays, polarised radio continuum and 100~$\mu$m emission \citep{Heiles:1999p15184,West:2007p15460,McClureGriffiths:2001p13769}. We find no convincing evidence of soft X-rays \citep{Snowden:1995p15529} (due to confusion) or 1420 MHz polarised emission \citep{Testori:2008p15690} associated with the shell. We also looked at the H$\alpha$ data from the Southern H-Alpha Sky Survey Atlas \citep[SHASSA;][]{Gaustad:2001p19243} H$\alpha$ imaging survey compiled by \citet{Finkbeiner:2003p15462}. Based on extinction \citep{Schlegel:1998p19628} and optical depth estimates \citep{Madsen:2005p19717} in this region of the Galaxy,  we can place an upper limit on the intensity of any possible H$\alpha$ emission by noting that we see no emission within the main shell structure ($b \sim -20^{\circ}$) or along the shell walls ($b \sim -25^{\circ}$) at limits of $[1.2 \pm 0.5, 2.0 \pm 0.5]$~R\footnote{One Rayleigh (R) $ = 10^6 / 4 \pi \textrm{ photons s}^{-1} \textrm{ cm}^{-2} \textrm{ sr}^{-1}$.} respectively, which, given estimated optical depths of $\tau = [0.23,0.46]$ at 656.26 nm, corresponds to maximum surface brightnesses of $[1.5 \pm 0.6, 3.1 \pm 0.8]$~R. 

At higher latitudes where the shell may be fragmenting, we note an evident H$\alpha$ structure in the SHASSA data at $l \sim 8^\circ, b \sim -30^\circ$ with an angular extent of $\sim 2^\circ$ and a peak brightness of $\sim 15$ R. This structure is on the interior of the fragmentary H{\sc i} emission seen at high latitudes (described in Section \ref{morph}). We find that the H{\sc i} at lower velocities correlates well with IRIS 100~$\mu$m emission \citep{MivilleDeschenes:2005p15697}. Fig. \ref{fg:halpha} shows the emission in H{\sc i}, H$\alpha$ and 100~$\mu$m at different velocities, showing correlation between the shell and the H$\alpha$ feature. To verify that the morphological association of the H$\alpha$ and H{\sc i} corresponds to physical association, we obtained an H$\alpha$ spectrum with the Wisconsin H-Alpha Mapper \citep[WHAM;][]{Haffner:2003p20906,Haffner:2010p21381} towards the region of the cloud with the brightest H$\alpha$ emission in the SHASSA data, $(l, b) = (8.70, -29.85)$. The data were obtained from Cerro Tololo Interamerican Observatory on the 16th November 2011. WHAM integrates all emission within its $1^\circ$-diameter beam. We also obtained an off-source H$\alpha$ spectrum centred at $(7.03, -33.92)$. We fit a single Gaussian modelling the geocoronal H$\alpha$ line to both the on- and off-source spectra and then subtracted the geocorona-subtracted off spectrum from the geocorona-subtracted on spectrum to remove the faint, mostly-unidentified atmospheric lines that dominate the baseline at the surface brightness sensitivity of WHAM \citep{Hausen:2002p21375}. We also used the geocoronal line to calibrate the velocity scale to an accuracy of less than $0.5 \textrm{ km s}^{-1}$ \citep{Haffner:2003p20906}. The resulting spectrum consists of entirely astronomical emission and is shown in Fig.~\ref{halphaspec}. The emission is well-fit by a single Gaussian component with a mean LSR velocity of $6.7 \textrm{ km s}^{-1}$, a full width at half-maximum of $27 \textrm{ km s}^{-1}$, and an intensity of $5$~R. Although the relatively large width of the H$\alpha$ line precludes definitive separation from local gas, the single-component Gaussian profile and the good agreement of the velocity centroid of the H$\alpha$ emission with the H{\sc i} data strongly support the association between the H$\alpha$ and the H{\sc i}-emitting gas which is suggested by the morphology from the SHASSA image.

While H$\alpha$ is observed in nearly all directions at high latitude \citep{Haffner:2003p20906}, multiple emission mechanisms are likely responsible in different directions. H$\alpha$ well off the Galactic disk is thought to be emitted either by gas which is photoionised in situ by ionising photons from OB stars in the disk which have escaped the plane \citep{Reynolds:1990p21452,Haffner:2009p20946,Wood:2010p21483} or to be H$\alpha$ emission from closer to the plane which has been scattered off of high-latitude dust clouds \citep{Witt:2010p20944,Seon:2011p20952}. H$\alpha$ can also be emitted shock-ionised gas or by gas photoionised by a harder (soft X-ray) radiation field such as might be produced by million-degree gas in the interior of the supershell. With future observations of the [O {\sc iii}] $\lambda 5007$ and [S {\sc ii}] $\lambda 6716$ emission lines in combination with photoionisation modelling \citep{Wood:2004p21484,Madsen:2006p21487}, we may be able to discriminate between emission mechanisms in this H$\alpha$ feature and determine how the H$\alpha$ relates to the supershell.

\section{H{\sc i} self-absorption in GSH~006$-$15+7}\label{hisa}
\begin{figure}
\centering
\includegraphics[angle=0,width= 0.48\textwidth]{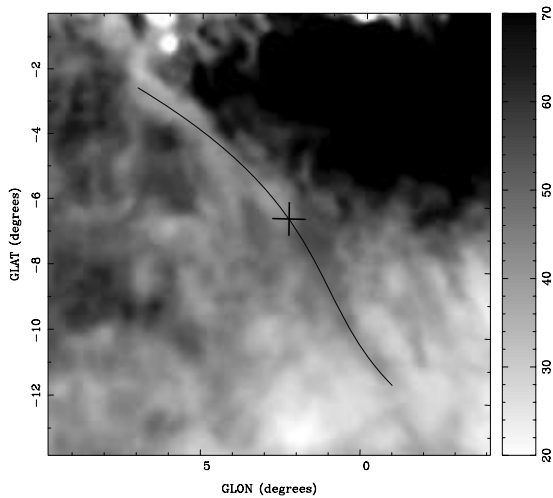}
\caption{Image of the region of the shell used for H{\sc i}SA analysis, at the systemic velocity of $\sim 7$ km s$^{-1}$. We scale the data linearly from 20 K to 70 K. The line traces the position of the shell wall and the cross marks the region where GSH~006$-$15+7 transitions from absorption to emission.}
\label{fg:hisa}
\end{figure}

\begin{figure}
\includegraphics[angle=0,width=0.5\textwidth]{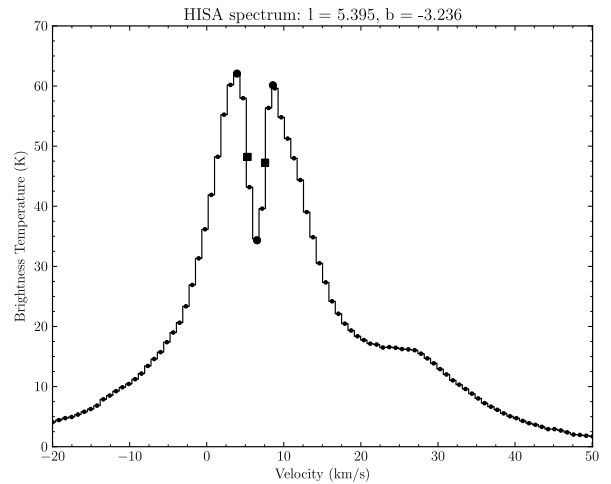}
\caption{Example spectrum at a single position ($l = 5.395, b = -3.236$) showing the absorption due to GSH~006$-$15+7. The maximum absorption is predominantly seen at the systemic velocity of 7 km s$^{-1}$ along the wall of the shell. The maxima/minima (circles) and estimated FWHM (squares) are shown overplotted on the spectrum and are used in Section \ref{hisacalc} to estimate a line-width of $\sim 2$ km s$^{-1}$ corresponding to a spin temperature of $\sim 40$ K.}
\label{fg:hisa2}
\end{figure}

A unique feature of GSH~006$-$15+7 is that we see the upper right shell wall transition clearly from H{\sc i} emission into absorption at $b \gtrsim -6.5^\circ$, with the deepest absorption occurring at the systemic velocity $v \sim 7$ km s$^{-1}$. Fig. \ref{fg:hisa} shows the transitioning wall at this systemic velocity, traced with a line. The region where we might expect to find the transition point is marked with a cross, showing a shift from emission to absorption. This phenomenon is known as H{\sc i} self-absorption, or H{\sc i}SA \citep{Heeschen:1955p18895,Gibson:2000p14405}. An absorption profile (such as the example spectrum shown in Fig. \ref{fg:hisa2}) can indicate the presence of cool atomic gas. We expect to see H{\sc i} absorption against a diffuse background $T_{\rm bg}$ that has a higher brightness temperature than the spin temperature $T_{\rm s}$ of the foreground gas. However, the detection of absorption is not generally enough to constrain the spin (or excitation) temperature of the gas and so it is usually assumed, constrained based on the presence of other molecules or estimated via line width fitting \citep{Gibson:2000p14405}. In special cases, a transition from H{\sc i} emission to absorption allows $T_{\rm s}$ to be measured directly at the transition point \citep{Kerton:2005p12562}. 

We apply  H{\sc i}SA diagnostics to this feature below in order to estimate spin temperature $T_{\rm s}$, optical depth $\tau$ and column density $N_{\rm H}$ in the shell. We note that there is some spatial proximity of GSH 006$-$15+7 to the local self-absorbing Riegel-Crutcher cloud at a distance of $\sim 100$~pc towards the Galactic centre \citep{McClureGriffiths:2006p20000}, with their angular separation varying from a few to about 15 degrees. Taking the point of maximum absorption to represent the systemic velocity of Riegel-Crutcher gives a peak absorption velocity of $\sim 5.8$ km s$^{-1}$, compared to the peak absorption velocity of GSH~006$-$15+7 at $\sim 7.2$ km s$^{-1}$. The two objects also show differing absorption profiles and are morphologically distinct, but with some evident overlapping absorption at latitudes $> 0^\circ$. In our H{\sc i}SA analysis, we use only latitudes $< 0^\circ$ in order to avoid any confusion. 

In a simplified case of absorption in the context of radiative transfer, we can define an `on-source' brightness temperature ($T_{\rm b}$) where we observe the absorption through the absorbing cloud (also known as the observed temperature, which in this case is the continuum-subtracted H{\sc i} brightness temperature), and an `off-source' brightness temperature ($T_{\rm bg}$) where we estimate the background without the absorption present. Under the limiting, but reasonable, assumptions that any background or foreground gas has negligible optical depth \citep{Gibson:2000p14405}, the difference between the on and off is

\begin{equation}\label{hisaeq0}
T_{\rm b} - T_{\rm bg} = (T_{\rm s} - T_{\rm c} - p T_{\rm bg})(1-e^{-\tau})
\end{equation}
where $T_{\rm s}$ is spin temperature of the absorbing cloud, $p$ is the fraction of H{\sc i} emission originating behind the H{\sc i}SA cloud (as opposed to in front of the cloud), and $T_{\rm c}$ is the mean continuum background \citep{Feldt:1993p14415}. A value of $p$ = 1 assumes that all H{\sc i} emission originates from behind an object seen in absorption. Although we know there is some foreground material towards GSH~006$-$15+7, setting $p$ = 1 is a reasonable assumption given the relatively small path length to the shell. In the case of a transition from emission to absorption, the transition point marks the place where the background $T_{\rm bg}$ and the source $T_{\rm b}$ are equal in magnitude; in this case Equation \ref{hisaeq0} simplifies to

\begin{equation}\label{hisaeq1}
T_{\rm s} = p T_{\rm bg} + T_{\rm c}. 
\end{equation}
Equation \ref{hisaeq1} thus allows direct calculation of the spin temperature under certain circumstances at the transition point in GSH~006$-$15+7, giving insight into the local properties of the H{\sc i} gas. Once we estimate the spin temperature $T_{\rm s}$, it is possible to substitute the values for $T_{{b}}$,  $T_{\rm bg}$, $T_{\rm s}$ and $T_{\rm c}$ to solve for optical depth, using 

\begin{equation}\label{hisaeq2}
\tau = -\textrm{ln} \left( 1 - \frac{T_{\rm b} - T_{\rm bg}}{T_{\rm s} - p T_{\rm bg} - T_{\rm c}} \right).
\end{equation}
Using Equation \ref{hisaeq2} along the wall of GSH~006$-$15+7 under the assumption of constant $T_{\rm s}$, we can calculate $\tau$ and use this to estimate the corresponding column density, using

\begin{equation}\label{hisaeq3}
N_{\rm H} = 1.823 \times 10^{18} T_{\rm s} \int \tau dv
\end{equation}
which, assuming a Gaussian form for the integral of $\tau(v)$,  simplifies to

\begin{equation}\label{hisaeq3}
N_{\rm H} = \frac{\tau_0 T_{\rm s} \Delta v}{C},
\end{equation}
where $N_{\rm H}$ is the column density, $\tau_0$ is the line centre opacity, $T_{\rm s}$ is the spin temperature, $\Delta v$ is the line profile width and $C$ is a standard correction term \citep{Gibson:2000p14405} [$5.2 \times 10^{-19}$ cm$^2$ K km s$^{-1}$] \citep{Dickey:1990p13348}. From here, we can redo our previous mass calculation under the assumption that this newly estimated column density better represents the true value as the unlikely assumption of thin optical depth is no longer made. 

\subsection{H{\sc i}SA results for GSH~006$-$15+7}\label{hisacalc}
\begin{figure}
\includegraphics[angle=0,width=0.5\textwidth]{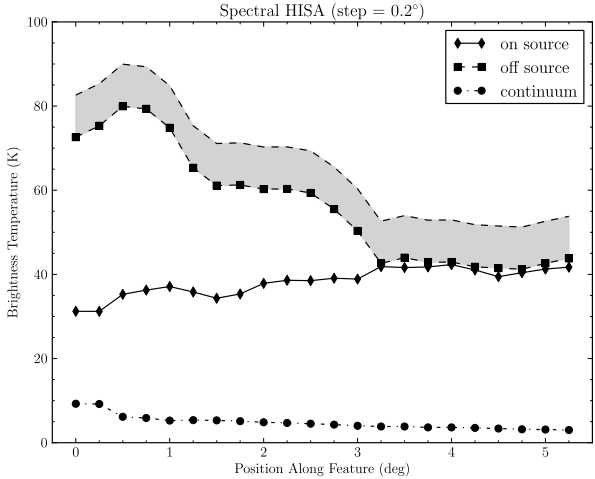}
\caption{Spectrally-derived variation in brightness temperature on/off the shell arm. This plots the maximum absorption at each latitude (diamonds) versus the background temperature (squares) and the continuum background (circles, Calabretta et al, in preparation), with the amount of absorption maximised along the wall of the shell. The horizontal axis is zeroed at $b = -2^\circ$. The grey-shaded region represents the extent to which we believe the background is uncertain ($\sim 10$ K), which is incorporated into the calculations. The transition point occurs where the curves meet, but the consequent emission of the shell is not clear due to the difficulty of spectrally separating emission due to GSH~006$-$15+7 and other emission.}
\label{fg:hisa4}
\end{figure}

We examine the absorption to emission transition in velocity-longitude space along the wall of the shell at 0.2$^\circ$ increments over the range $b = [-7.0, -2.0]^\circ$, obtaining the spectrum at the velocity of maximum absorption for each latitude point. We resample each spectrum in order to ease the location of minima and maxima along the curve, taking into account the velocity resolution of $\sim 1$ km s$^{-1}$ in estimating the velocity width of each profile. By obtaining the minimum and maximum points that define the absorption, we adopt these as our on-source ($T_{\rm b}$) and off-source ($T_{\rm bg}$) values. We determine the continuum background ($T_{\rm c}$) from H{\sc i} Parkes All Sky Survey (H{\sc i}PASS) continuum data at 1.4 GHz corrected for extended emission (Calabretta et al, in preparation). The results are plotted in Fig. \ref{fg:hisa4}, showing the on-source $T_{\rm b}$, the averaged off-source $T_{\rm bg}$ and the background continuum $T_{\rm c}$. 

We assume the simplest case for the background by taking the average of each pair of maxima to represent the amount to which the emission is absorbed. To investigate the error on our assumed background, we examined longitude profiles at 7 km s$^{-1}$ at the latitude of each point used in our analysis and find the peak brightness temperature on the right side of the spatially-apparent absorption (outside the shell wall). This allows us to check how well the background derived from the velocity spectra agrees with a spatially-determined background temperature. We find a positive trend indicating that the spectral profile does underestimate the background, with a peak deviation of $\sim$~10~K and a mean deviation of $\sim$~4~K.  We thus assume that our background estimate could be as much as $\sim 10$ K underestimated and factor this into our predictions below.

The spectral method we have used, although reliable, does not easily allow visualisation of the emission of GSH~006$-$15+7 after the transition point because it is difficult to separate the shell emission from background emission without stable background spectral profiles that can be subtracted. Due to this, we cannot use Equation \ref{hisaeq1} to tightly constrain the spin temperature at the transition point, but we can set an upper limit to the spin temperature by considering the combined $T_{\rm b}$ and $T_{\rm c}$ values along the wall,

\begin{equation}\label{hisaeq4}
T_{\rm s,max} = T_{\rm b} + T_{\rm c},
\end{equation}
as in our simple radiative transfer this will always be an overestimate as there are likely to be sources contributing to the emission along the line of sight that are not taken into account. We thus calculate $T_{\rm s,max}$ systematically along the shell wall and then average the results to give a mean maximum spin temperature of $\overline{T_{\mathrm s,max}} \sim 40$ K. Any fluctuations in spin temperature due to the local properties of the gas in the shell wall are likely to be moderated by using the mean value. To check the value of $T_s$, we estimate the width of the absorption profile in Fig. \ref{fg:hisa2} and find the FWHM $\sim 2$ km s$^{-1}$. In this case, the resulting thermal velocity dispersion $\sigma_v$ is $\textrm{FWHM}/2.36 \sim 0.8$ km s$^{-1}$ with a kinetic temperature of

\begin{equation}\label{hisaeq4}
T_{\rm k}  = \frac{m_{\rm H} \sigma_v^2}{k},
\end{equation}
where $m_{\rm H}$ is the mass of hydrogen and $k$ is Boltzmann's constant. This yields $T_{\rm k} \sim$ 90 K. For temperatures $T_{\rm k} < 1000$~K in the cold neutral medium (CNM), $T_{\rm k}$ is essentially equal to $T_{\rm s}$ \citep{Liszt:2001p19905,Roy:2006p19909}. However, spectral modelling of H{\sc i} absorption and emission profiles demonstrates that significant line broadening relative to the spin temperature at peak opacity $\tau_{max}$ is generally seen \citep{Heiles:2003p19192} and can indicate either or both of significant internal temperature structure in the gas and a turbulent contribution to the line-width measurement. Our spin temperature estimate of 40 K implies excess line-broadening of $\sigma_v \sim$ 0.6 km s$^{-1}$, which is fully consistent with values typically found in the ISM.

Using Equation \ref{hisaeq2} along the absorbing wall of GSH~006$-$15+7 and a spin temperature of 40 K gives values of $\tau$ in the range [1.8, 5.0] with a mean of $\tau \sim 3.1$. The potential underestimation of the background of up to $\sim 10$ K is used to calculate the upper limit in the given ranges. We use the optical depths to estimate the corresponding column density, which ranges over [2.7,~7.7]~$\times 10^{20}$~cm$^{-2}$ with a mean value of $N_{\rm H} \sim 4.7 \times 10^{20}$~cm$^{-2}$. We compare this to the value estimated from the zeroth-moment map in Section \ref{mass} of $N_{\rm H} \sim 2 \times 10^{20}$~cm$^{-2}$, and note that the mean column density is more than double the previous estimate, which is to be expected due to eliminating the assumption of negligible optical depth. 

If we assume that our new estimate better reflects the true column density and substitute this into the calculation for mass, we find that $M$ ranges over [1.8, 5.1] $\times 10^6~{\rm M_\odot}$ with a corresponding number density of [1.7, 4.8] atoms cm$^{-3}$. This is in comparison to our previously determined estimate of $\sim 10^6~{\rm M_\odot}$ and $\sim 1.3$ atom cm$^{-3}$, where the assumption of zero optical depth leads to a significant underestimation of the shell mass. We note that the application of a single opacity correction factor to the entire supershell may be erroneous as it would be plausible to have systematically lower densities and opacities at larger off-plane distances, while the opacity measurement is done closer to the plane where these would be maximised. To counter the effect of this, we use the range of resultant masses from using optically thin versus H{\sc I}SA as a representation of the error on the mass such that $M = 3 \pm 2 \times 10^6~{\rm M_\odot}$. 

\section{Evolution of GSH~006$-$15+7}\label{model}
To investigate further the nature of the shell, we have searched for potential stellar powering sources and applied simple modelling in order to constrain the origin of GSH~006$-$15+7. This involves estimating the amount of energy deposited by candidate sources through stellar winds and supernovae in order to assess whether these regions could form GSH~006$-$15+7 based on the energy predictions we made in Section \ref{mass}, where we found $N_* \sim 26$ and $E \sim 10^{52}$ ergs. 

\subsection{Energy estimation and supernovae}\label{soureg}
\begin{figure}
\includegraphics[angle=-90,width=0.48\textwidth]{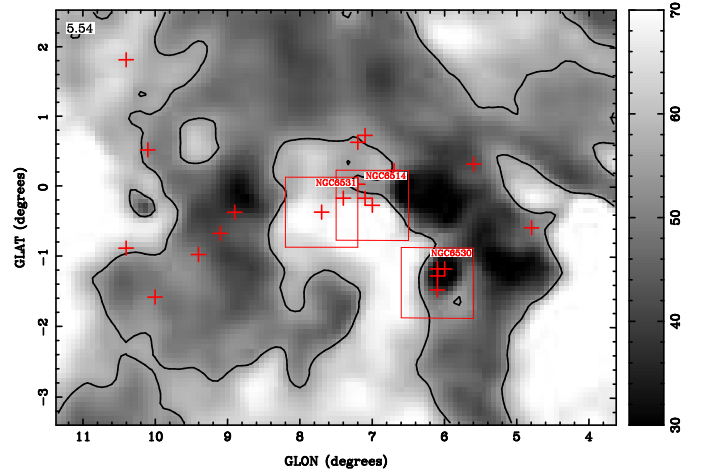}
\caption{The assumed source region of GSH~006$-$15+7, showing the stars of Sagittarius OB 1 (red crosses) and the open clusters (red boxes) used for stellar wind estimates and for estimating the energy contribution of supernovae overlaid on a H{\sc i} velocity slice at $\sim 5$ km s$^{-1}$ (where the cavity these objects occupy is most clearly visible). The black contour is at 55 K to show the outline of the shell and the scale is linear from 30 K to 70 K. }
\label{fg:clust}
\end{figure}

We use multiwavelength data and published databases to investigate the region of the Galactic plane where the walls of GSH~006$-$15+7 meet to search for powering sources. Of the nearby OB associations \citep{Humphreys:1978p12865}, Sagittarius (Sgr) OB~1 is a likely candidate based on coordinates as well as its reported distance of 1.58 kpc. The 13 high-mass stars of Sgr~OB~1 are spread over 10$^\circ$ of longitude, which makes it difficult to assess their true association. At the reported distance of $\sim 1.6$ kpc, the mean longitudinal drift of each star from the centre of the Sgr OB 1 association corresponds to roughly 40 pc, in agreement with an average stellar drift velocity of $\sim 5$~km s$^{-1}$ \citep{McCray:1987p12912} and an age of around 10$-$20~Myr for GSH~006$-$15+7. We show their distribution in Fig. \ref{fg:clust}, as well as the distribution of the nearby and possibly associated open clusters NGC 6514, NGC 6530, and NGC 6531 (with which some of the stars of Sgr~OB~1 are associated). The general area has also been referred to as Sagittarius I \citep{Stalbovsky:1981p13958}, incorporating a number of clusters and nebulae. Overall, the entire region shows evidence of high star formation activity, and so we base our initial energy calculations on the stellar winds of the known high-mass stars. 

We calculate stellar wind estimates using Sgr OB 1, of which stellar types are well known \citep{Mason:1998p18137,MaizApellaniz:2004p18142} and can be used to determine effective temperature, luminosity, mass, maximum age, mass loss rate, wind terminal velocity and wind luminosity. We estimate the stellar wind mechanical luminosity as $L_{\rm SW} = \frac{1}{2}\dot{M} V_\infty^2$, based on the observed spectral type which allows derivation of mass loss rate $\dot{M}$ and wind terminal velocity $V_\infty$. We match the stellar types earlier than B0 to those for the Carina Nebula \citep{Smith:2006p18022} and account for luminosity classes II and IV (which are generally not included in models) by raising them to the next highest class (i.e. I and III), which provides an overestimate for their energy contribution. We have also assumed that the age of each star is the lifetime of a star with its estimated mass, which gives us an upper limit. Given the resulting wind luminosity, it is then possible to estimate total energy input from these stars due to their stellar winds.

It is expected in studies of large old shells that many of their forming high-mass stars will have exploded as supernovae (SNe), as their lifetimes are on the order of a few to several million years \citep{Chiosi:1992p19813}. As such the energy calculated from stellar winds alone of any powering region is likely to underestimate its true energy contribution, and so we assess the contribution of stars that have already gone supernova\footnote{We refer to these stars as exploded stars.}. We can estimate the number of supernovae based on the initial mass function (IMF), a measure of mass distribution. This can be readily done in the case of stellar clusters with a known IMF, as opposed to OB associations whose stars can be scattered across large regions of space and possibly even associated with different stellar clusters (as in the case of Sgr OB 1). We perform a cursory IMF calculation for Sgr OB 1 but emphasise that there are strong assumptions impacting this estimate. For this reason, we focus the IMF calculation on the open clusters NGC 6514, NGC 6530 and NGC 6531 due to their proximity to GSH~006$-$15+7, distance close to 1.5 kpc and stellar content, noting that several of the stars of Sgr OB 1 are associated with these clusters. We do not include the other positionally nearby clusters NGC 6546 (due to its large estimated age exceeding the shell's age) or Bochum 14 (due to its lack of high-mass stars and poorly known stellar composition). 

For the clusters, we use UBV photometry and stellar classification of members to calculate effective temperature (log $T_{\rm eff}$), bolometric correction ($BC$) and V-band extinction ($A_{\rm V}$) of each star \citep{Massey:1995p17301}, using these values to then calculate the absolute magnitude ($M_{\rm V}$), bolometric magnitude ($M_{\rm bol}$) and hence luminosity (log $L/{\rm L}_\odot$) \citep{Crowther:2004p17326}. Masses are estimated using this resulting luminosity via the mass-luminosity relationship \citep{Eddington:1924p19538}, with awareness that this relation is generally valid only for main sequence stars. We make the assumption that the photometry is complete from the minimum high-mass star calculated ($M_{\rm min}$) to the maximum high-mass star calculated ($M_{\rm max}$), with any stars higher than this mass already gone supernova. We adopt a minimum possible stellar mass of 0.5~${\rm M_\odot}$ \citep{Kroupa:2001p17568} and a maximum possible stellar mass of 150~${\rm M_\odot}$ \citep{Figer:2005p17553}, and integrate over the range assumed complete

\begin{equation}\label{vesc}
F_{\rm N}(M_{\rm min} < M < M_{\rm max}) = \frac{\int^{M_{\rm max}}_{M_{\rm min}} M^{\alpha} {\rm d}M}{\int^{150}_{0.5}  M^{\alpha} {\rm d}M},
\end{equation}
where $F_{\rm N}$ is the number fraction of stars in the given range with respect to the total and $\alpha$ is the slope of the initial mass function. We can therefore estimate the total number of cluster stars by dividing the number of stars $N$ in this range by $F_{\rm N}$. Then, we estimate the number fraction of exploded stars in the cluster,
 
 \begin{equation}\label{vesc}
F_{\rm N}(M > M_{\rm max}) = \frac{\int_{M_{\rm max}}^{150} M^{\alpha} {\rm d}M}{\int^{150}_{0.5}  M^{\alpha} {\rm d}M}, 
\end{equation}
and multiply by the total number of stars to obtain an estimate of the total number of exploded stars which we attribute to supernovae \citep{Bruhweiler:1980p15707,Heiles:1987p19728}.  We assume that each exploded star has contributed $10^{51}$ ergs of energy through stellar winds over its lifetime and $10^{51}$ ergs via its supernova, giving an energy contribution of $2 \times 10^{51}$ ergs per exploded star.

\subsection{Cluster parameters}
We obtain UBV photometry and stellar types for each cluster from the online {\sc webda} (Web Open Cluster Database \citep{Mermilliod:2003p19501}. As we are necessarily assuming that these clusters contributed to the formation of GSH~006$-$15+7, we adopt a distance of 1.5 kpc to all clusters. We take the age estimates from the catalogue of \citet{Dias:2002p17203}, and obtain values of mean reddening from the literature. We obtain values for the IMF slope ($\alpha$) of each cluster where available, otherwise we adopt the traditional Salpeter value of $\alpha = -2.35$ \citep{Salpeter:1955p16905}. We use in our calculations a total of 60 stars from the three clusters described below.

\begin{table*}
\begin{tabular}{l c c c c c c c}
\hline
Cluster $^{(1)}$ & M $> 8$ M$_\odot$ $^{(2)}$  & M$_{\rm min}$ (M$_\odot$) $^{(3)}$  & M$_{\rm max}$ (M$_\odot$) $^{(4)}$  & $\alpha$ $^{(5)}$  & Cluster stars $^{(6)}$  & \# SNe $^{(7)}$ & Energy $^{(8)}$\\ 
\hline
NGC 6514 & [4, 11] & [5, 7] & [23, 65] & $-2.35$ & [360, 520] & [2, 1] & [4.0, 2.0] $\times 10^{51}$ ergs\\
NGC 6530 & [20, 30] & [5, 8] & [28, 83] & $-2.22$ & [590, 970] & [4, 1] & [8.0, 2.0] $\times 10^{51}$ ergs \\
NGC 6531 & [5, 9] & [4, 6] & [17, 42] & $-2.29$ & [260, 400] & [3, 2] & [6.0, 4.0] $\times 10^{51}$ ergs \\
Total & ... & ... & ... & ... & ... &[9, 4] &[1.8, 0.8] $\times 10^{52}$ ergs \\ \hline
Sgr OB 1 & [13, 13] & [17, 19] & [43, 56] & $-2.35$ & [2100, 2300] & [4, 3] &[1.0, 0.6] $\times 10^{52}$ ergs
\\

\hline 
\end{tabular}~\\
\caption{Parameters of each cluster used to model IMF and resulting values. The columns are: (1) cluster name, (2) number of stars greater than 8 solar masses, (3) minimum mass used, (4) maximum mass used, (5) initial mass function slope, (6) extrapolated number of cluster stars based on IMF predicted number fraction, (7) number of supernovae estimated based on the mass distribution and (8) corresponding energy of these stars, assuming 10$^{51}$ ergs from stellar winds and 10$^{51}$ ergs from the supernova. In all cases for the clusters, the first number represents a mass power of 4, and the second number represents a mass power of 3. In the case of Sgr OB 1, the numbers represent masses from different sources \citep{Martins:2005p18681,Weidner:2010p18299}.}\label{tb:imf}
\end{table*} 

\subsubsection{NGC 6514}
This object is also known as the Trifid Nebula or M20, housing an open cluster of around 360 known stars as well as various types of nebulae and dense regions of dust. The distances reported for this object are quite varied, from 0.8 kpc to over 2 kpc, however we adopt a distance of 1.5 kpc, the Dias catalogue age of 23.3 Myr and a mean reddening of $E(B-V)$ = 0.23 \citep{Ogura:1975p17101}. We use an IMF slope of $\alpha = -2.35$ for this cluster.

\subsubsection{NGC 6530}
NGC 6530 is a well-studied young, active star-forming cluster associated with the Lagoon Nebula, described as the core cluster of the Sgr OB 1 association \citep{Chen:2007p17014} and housing roughly 350 known stars. Of the three clusters, it is the most active and also houses the highest number of O stars. Based on the literature, we adopt a mean reddening of $E(B-V)$ = 0.35 \citep{Sung:2000p17013} for this cluster, the Dias catalogue age of 7.4 Myr and an IMF slope of $\alpha = -2.22$ \citep{Prisinzano:2005p12624}. 

\subsubsection{NGC 6531}
This cluster is not particularly well studied due to its small number of detected stars (around 100), lack of numerous high-mass stars, nebulae or bright star forming regions. It has reported distance moduli of 11.35 \citep{Mermilliod:1986p15235}, 10.70 \citep{Forbes:1996p13948}, 10.5 \citep{Park:2001p13951} and most recently 10.47 \citep{McSwain:2005p13956}, giving a distance range of 1.2 to 1.9 kpc with a mean of 1.4 kpc. Its age has been estimated with a reported range of 8$-$11~Myr. We adopt the Dias catalogue age of 11.7~Myr, a mean reddening of $E(B-V)$ = 0.28 for this object and an IMF slope of $\alpha = -2.29$ \citep{Forbes:1996p13948}. 

\subsection{Modelling the energy of the shell: stellar winds}\label{winds}
In the case of Sgr OB 1, we estimate~$\sim 3 \times 10^{51}$~ergs of stellar wind energy, too low to account solely for the formation of GSH~006$-$15+7. Similarly we obtain a total cluster stellar wind energy of $\sim 1 \times 10^{51}$ergs, which we expect because the clusters have fewer high-mass stars overall. As only 20 per cent of these winds would have contributed to the expansion of the shell and our energy prediction was 10$^{52}$ ergs, we conclude that the stellar winds of Sgr~OB~1 or the clusters could not be the sole powering source of GSH~006$-$15+7, which is not necessarily surprising given the size and age of the shell. We therefore assume that the majority of the expansion energy must have come from supernovae that are now invisible.

\subsection{Modelling the energy of the shell: supernovae}
We now estimate the amount of energy we could derive based on exploded stars from the clusters we have examined. As before we adopt a distance of 1.5 kpc, values of reddening from the literature and the \citet{Dias:2002p17203} catalogue cluster age estimate. The results are summarised in Table~\ref{tb:imf}. We obtain a maximum energy input from supernovae and stellar winds (from exploded stars) of around $\sim 2 \times 10^{52}$ ergs total. But using the same conversion efficiency of supernova energy used in Section \ref{mass} of 20 per cent gives us only $E_{\rm SNe} \sim 5 \times 10^{51}$ ergs.

For comparison, we perform the same calculation for the stars of Sgr~OB~1 (for which we have more reliable masses) but emphasise that this approach implicitly assumes that the association stars formed from the same mass distribution. This is likely to be an over-simplification. We adopt mass estimates from two sources \citep{Martins:2005p18681,Weidner:2010p18299} and use this to give a likely energy range. We use the standard Salpeter value of $\alpha = -2.35$ for the IMF slope and assume that the association stars are complete from the minimum detected mass (17 M$_\odot$/19 M$_\odot$) to the maximum detected mass (43 M$_\odot$/56 M$_\odot$), where the two numbers represent mass estimates from different sources \citep{Martins:2005p18681,Weidner:2010p18299}. This results in an energy range of 0.6$-$1.0~$\times 10^{52}$ ergs total, which is essentially in agreement with the cluster estimate as suggesting a maximum converted energy on order $\sim 2 \times 10^{51}$~ergs. 

We find here that the formation of GSH~006$-$15+7 cannot be accounted for solely by the supernovae and stellar winds of exploded stars from either the clusters or Sgr~OB~1. The combination of stellar wind energy found in Section \ref{winds} ($E_{\rm wind} \sim 5 \times 10^{50}$ ergs) with that of the exploded stars ($E_{\rm SNe} \sim 5 \times 10^{51}$ ergs) is close to $10^{52}$~ergs (with the dominant contribution being from exploded stars). We also expect that our rough assumption of $2 \times 10^{51}$ ergs per exploded star may in fact be an underestimate of the true energy contribution of the very high-mass stars, since the winds are likely to exceed the supernova energy for high-mass stars earlier than O6 \citep{Abbott:1982p20871}. It therefore seems reasonable to assume that the shell was formed by the stars in and or around the Sgr OB 1 association.

\section{Conclusion}\label{disc}
We have reported on the discovery, properties and analysis of a new Milky Way supershell GSH~006$-$15+7. This object spans longitudes of $l = [356, 16]^\circ$  and latitudes of $b = [-28, 2]^\circ$, which translates into physical dimensions of $(790\pm260) \times (520\pm175)$~pc at an estimated distance of $1.5 \pm 0.5$~kpc. The shell is elongated in the latitude direction, which is possibly due to a density gradient away from the Galactic plane. We estimate the dynamical age to be between 7$-$15~Myr, with the coherent structure of the shell suggesting an age less than 20~Myr based on known supershells in the Galaxy. Energy estimates suggest a formation energy most likely $\sim 10^{52}$~ergs, with a mass estimate, corrected for optical depth, of $3 \pm 2 \times 10^6$~M$_\odot$. 

We have found no convincing evidence of an upper shell counterpart. In the case of an absent upper shell complementing GSH~006$-$15+7, we can interpret this as the result of an asymmetric outflow, due possibly to inhomogeneous gas surrounding the source or a source region slightly below the Galactic plane which preferentially outflows in the direction of the lowest density. An offset of even 100 pc from the Galactic plane, depending on the vertical density distribution, is known to have a significant effect on the direction of outflow and shape of the resulting supershell \citep{Chevalier:1974p20157}.

We have investigated a self-absorbing feature along the shell wall in order to place a more realistic constraint on our estimates of shell mass as well as giving an indication of gas temperature. We have found that the gas in the supershell is comparatively cool with a spin temperature of $\sim 40$ K as well as being quite optically thick (and hence dense) with $\tau \sim 3$ which is supportive of recent studies showing correlation between supershells and molecular gas formation, which requires dense cool gas \citep{Dawson:2008p12949,Fukui:2010p20762,Dawson:2011p17008}.  

We see strong evidence of past star formation in the region where we would expect to find a powering source based on the morphology and physical properties of the supershell. We also find some evidence in favour of a powering source for GSH~006$-$15+7 via the OB association Sgr OB 1 and the nearby open clusters NGC 6514, 6530 and 6531. In analysing the energy contribution from stellar winds from existing stars and the winds/supernovae of exploded stars from Sgr OB 1 and the clusters, the estimated combined $\sim 5 \times 10^{51}$ ergs energy is close to the estimated requirement of $\sim 10^{52}$ ergs of formation energy, and is within the errors in our analysis. The absence of presently observable high-mass stars contributing significant stellar winds suggests an expansion partially, if not mostly, driven by the winds and supernovae of exploded stars, which likely originated as part of the same region occupied by Sgr OB 1 and the clusters.

Based on our analysis, we suggest that GSH~006$-$15+7 is in the transition stage between supershell and chimney. The filamentary structures we see in the morphology of the shell at high latitudes, evident in H{\sc i} and IR with possible associated H$\alpha$ emission, may be an indication that fragmentation and blowout is already beginning to take place. 

\section*{Acknowledgments} 
We are grateful for the use of 21 cm continuum data from the H{\sc i}PASS survey (Calabretta et al., 2011, in preparation) for the H{\sc i}SA background calculation. This research has made use of the {\sc simbad} database, the {\sc vizier} catalogue access tool and {\sc aladin}, operated at CDS Strasbourg France and the {\sc webda} tool, operated at the Institute for Astronomy of the University of Vienna. WHAM and SHASSA are each supported by the US National Science Foundation. Smooth remote operations of WHAM are made possible by L. M. Haffner, K. A. Barger, and K. P. Jaehnig at the University of Wisconsin-Madison and the excellent staff at the Cerro Tololo Interamerican Observatory, particularly O. Saa. We also would like to thank J.R. Dawson, T. Murphy, F.J. Lockman, K.J. Brooks, J.E.G. Peek, J.A. Green, S.L. Breen, J.M. Rathborne, T. Robishaw, S.A. Farrell, M.L.P. Gunawardhana and G.A. Rees for helpful discussions during this research. 

\bibliography{all}{}
\bibliographystyle{mn2e}

\end{document}